\documentstyle[preprint,pre,aps,eqsecnum]{revtex}
\draft
\begin{document}
\draft
\title{\bf REGULARISATION OF FUNCTIONAL \\
DETERMINANTS USING BOUNDARY PERTURBATIONS}
\author{Alan J. McKane$^{\dag}$ and Martin B. Tarlie$^{\ddag}$}
\address{$^{\dag}$Department of Theoretical Physics \\
University of Manchester \\ Manchester M13 9PL, UK \\
$^{\ddag}$Department of Physics, 1110 West Green Street \\
and Materials Research Laboratory, 104 South Goodwin Avenue \\
University of Illinois at Urbana-Champaign \\
Urbana, Illinois 61801, USA}
\date{\today}
\maketitle
\begin{abstract}
The formalism which has been developed to give general expressions
for the determinants of differential operators is extended to the physically
interesting situation where these operators have a zero mode which has been
extracted. In the approach adopted here, this mode is removed by a novel
regularisation procedure, which allows remarkably simple expressions for these
determinants to be derived.
\end{abstract}
\section{Introduction}

The increasing use of path integrals as a calculational tool has led to a
corresponding increase in interest in the evaluation of functional
determinants. This is simply because the evaluation of Gaussian path integrals
typically gives such determinants. The first results were obtained over thirty
years ago: Gel'fand and Yaglom \cite{gy} derived expressions for the functional
determinants obtained from evaluating path integrals with the simplest type
of quadratic action. In subsequent years the results became more general and
the formalism more elaborate \cite{ls,dd}, culminating with the work of
Forman \cite{For} who gave a remarkably simple prescription which can be
applied to a rather general operator and boundary conditions.

However, in many calculations involving Gaussian integrals which are
currently carried out, these results are not directly applicable. The reason
is that the Gaussian nature of the integral is frequently a consequence of
expanding about some non-trivial \lq\lq classical" solution of the model
(e.g. a soliton or instanton). Typically this results in a particular point
(in space or time) being selected, which breaks the translational invariance
of the theory, and so gives rise to a Goldstone mode. There are other
possible ways that such a zero mode could come about, but in all cases the
Gaussian approximation breaks down. The remedy is to first extract this mode
as a collective coordinate \cite{Raj} and to treat only the non-zero modes in
the Gaussian approximation. Therefore, it is not the functional determinant
which is required in these cases --- it will in any case be identically
zero --- but the functional determinant with the zero mode extracted.

In this paper we present a systematic method to calculate this quantity.
The most obvious way to proceed is to \lq\lq regularise" the theory in some
way, so that the eigenvalue of the operator under consideration which was
previously zero is now non-zero. The determinant is now also non-zero and the
pseudo-zero eigenvalue can be factored out, the regularisation removed, and a
finite result obtained. Previous approaches have been rather ad hoc, being
performed on a case by case basis as the need arose. For example, it may be
possible in certain cases to modify the form of the operator in such a way that
the zero mode is regularised, but also that the calculation may still be
performed \cite{blz,tsg}.
In other cases, it may be possible to move the boundaries
to achieve the same end \cite{bbm,dku}.
Here we adopt an approach which applies to
very general situations and which, we believe, is the simplest and most
systematic regularisation and calculational procedure. This is because the
method is the least intrusive --- the operator and the position of the
boundaries are left unchanged --- only the form of the boundary conditions are
modified in the regularisation procedure. We use the notation and general
approach of Forman to calculate the regularised functional determinant, since
it is ideally suited to this form of regularisation, emphasising as it does
the separation of the boundary conditions from the solutions of a homogeneous
differential equation.

The outline of the paper is as follows. In section 2 we develop our method in
one of the simplest situations in order to clearly illustrate it. The
calculation of the regularised expression for the formerly zero eigenvalue is
derived, for the most general case that will interest us, in section 3 and
the general procedure for finding the functional determinant with the zero
mode extracted is described in section 4. In section 5 we apply the method to
certain specific cases and we conclude in section 6 with some general remarks.

\section{A simple example}

In this section we will explain the method by carrying out an explicit
calculation on what is perhaps the simplest example. Suppose that we wish
to calculate the determinant of an operator of the form
\begin{equation}
\label{simel}
L = \frac{d^2\,}{dt^2} + P(t) \ \ \ \ \ t \in [a,b]
\end{equation}
where $P(t)$ is a known real function. We suppose that the boundary conditions
on the functions on which $L$ operates is $u(a)=u(b)=0$. In particular, the
eigenfunctions of $L$ have to satisfy these conditions.

We now give Forman's prescription for calculating $\det L$. A more detailed
discussion is given in section 4, where our approach is explained in greater
generality. The recipe has two ingredients:

(i) Write the boundary conditions on $L$ in the form
\begin{eqnarray}
\label{defmn}
M\left[\matrix{u(a)\cr
               \dot{u}(a)\cr}\right] \ \ +
N\left[\matrix{u(b)\cr
               \dot{u}(b)\cr}\right] \ \ =
\left[\matrix{0\cr
              0\cr}\right]
\end{eqnarray}
where $M$ and $N$ are $2\times 2$ matrices and $\dot{u}=du/dt$. These two
matrices are not unique; for the case of our boundary conditions $u(a)=u(b)=0$
we choose them to be
\begin{eqnarray}
\label{simmn}
M=\left[\matrix{1&0\cr
                0&0\cr}\right], \ \ \
N=\left[\matrix{0&0\cr
                1&0\cr}\right].
\end{eqnarray}

(ii) Now consider a different problem. Let $y_1 (t)$ and $y_2 (t)$ be two
independent solutions of the homogeneous differential equation $Lh=0$.
Construct
\begin{eqnarray}
\label{hoft}
H(t)=\left[\matrix{y_1 (t)&y_2 (t)\cr
                   \dot{y}_1 (t)&\dot{y}_2 (t)\cr}\right],
\end{eqnarray}
and the $2 \times 2$ matrix $Y(b)\equiv H(b)H^{-1}(a)$.

Forman then proves that \cite{For}
\begin{equation}
\label{simres}
\frac{\det L}{\det \hat{L}} = \frac{\det (M+NY(b))}{\det (M+N\hat{Y}(b))}
\end{equation}
We would expect that $\det L$ itself is divergent, being a product of an
infinite number of eigenvalues of increasing magnitude. Therefore it is only
when it is defined relative to the determinant of an operator of a similar
type (denoted here by $\hat{L}$), that it has any meaning. In applications
to path integrals, ratios of determinants such as the one on the left-hand-side
of (\ref{simres}) naturally arise from the normalisation of the path integral
itself. In general, they will relate to a simple quantum system or stochastic
process, such as the harmonic oscillator or Ornstein-Uhlenbeck process. In
these cases, $\hat{P}(t)$ is independent of $t$ and will not, in general, have
a zero mode.

For the matrices $M$ and $N$ of our simple example,
\begin{eqnarray}
\label{simdetmn}
\det (M+NY(b)) & = & Y_{12}(b) \nonumber \\
& = & \frac{y_1 (a)y_2 (b) - y_2 (a)y_1 (b)}{y_1 (a)\dot{y}_2 (a) -
y_2 (a)\dot{y}_1 (a)}
\end{eqnarray}
The denominator of this expression is the Wronskian, which does not vanish
since the two solutions $y_1 (t)$ and $y_2 (t)$ are presumed independent. If
we take $y_1 (t)$ to be a solution for which $y_1 (a)=0$, then (\ref{simdetmn})
can be simplified to $y_1 (b)/\dot{y}_1 (a)$, so that, if $\hat{y}_1 (a)=0$
too,
\begin{equation}
\label{simdelL}
\frac{\det L}{\det \hat{L}} = \frac{y_1 (b)\dot{\hat{y}}_1 (b)}{\dot{y}_1 (b)
\hat{y}_1 (b)}
\end{equation}
This simple expression is particularly useful, since it only involves
$y_1 ,\hat{y}_1 $ and their first derivatives at one of the boundaries. We
should stress that results such as this have been known since the work of
Gel'fand and Yaglom --- our purpose here is to introduce the formalism
required to describe our approach, in as simple a way as possible.

Now suppose that $y_1 (b)=0$ (as well as $y_1 (a)=0$). Then $y_1 (t)$ is an
eigenvalue of $L$ with zero eigenvalue. This is the situation of interest to us
in this paper. To extract this zero mode, we first regularise the problem by
modifying it so that the operator is unchanged, but the boundary conditions
$u(a)=u(b)=0$ become
\begin{equation}
\label{simbc}
u^{(\epsilon )}(a)=0, \ \ \
u^{(\epsilon )}(b)=\epsilon \dot{u}^{(\epsilon )}(b)
\end{equation}
where $\epsilon$ is some small number. So now $y_1 (t)$ is no longer an
eigenfunction of $L$ with zero eigenvalue. Let $y^{(\epsilon )}_1 (t)$ be
the corresponding eigenfunction (i.e. the one which reduces to $y_1 (t)$ when
$\epsilon \rightarrow 0$) and let it have eigenvalue $\lambda ^{(\epsilon )}$.
To find $\det L$ with these boundary conditions we first note that $Y(b)$ is
unchanged, since it does not involve boundary conditions at all; it only
depends on two independent solutions to the homogeneous differential
equation $Lh=0$. Modifying the boundary conditions as in (\ref{simbc}) only
changes $M$ and $N$ to
\begin{eqnarray}
\label{simregmn}
M^{(\epsilon )}=\left[\matrix{1&0\cr
                0&0\cr}\right], \ \ \
N^{(\epsilon )}=\left[\matrix{0&0\cr
                1&-\epsilon\cr}\right].
\end{eqnarray}
This gives
$\det (M^{(\epsilon )}+N^{(\epsilon )}Y(b)) = Y_{12}(b)-\epsilon Y_{22}(b)$.
But since $y_1 (a)=y_1 (b)=0$, $Y_{12}(b)=0$, and so
\begin{eqnarray}
\label{simreg}
\det (M^{(\epsilon )}+N^{(\epsilon )}Y(b)) & = & -\epsilon Y_{22}(b)
\nonumber \\
& = & -\epsilon \frac{\dot{y}_1 (b)}{\dot{y}_1 (a)}
\end{eqnarray}
This is the regularised form of the determinant. In the next section we will
give a general method for finding $\lambda ^{(\epsilon )}$. In this simple
problem it turns out that, to lowest order,
\begin{equation}
\label{simlam}
\lambda ^{(\epsilon )} = -\epsilon \frac{\dot{y}^2_1 (b)}
{\langle y_1 | y_1 \rangle }
\end{equation}
where $\langle y_1 | y_1 \rangle$ is the norm of the zero mode:
\begin{equation}
\label{norm}
\langle y_1 | y_1 \rangle = \int^b_a dt \, y^2_1 (t) .
\end{equation}
{}From (\ref{simreg}) and (\ref{simlam}) we have that
\begin{equation}
\label{simdetpr}
\lim_{\epsilon \rightarrow 0}
\frac{\det (M^{(\epsilon )}+N^{(\epsilon )}Y(b))}{\lambda ^{(\epsilon )}} =
\frac{\langle y_1 | y_1 \rangle}{\dot{y}_1 (a)\dot{y}_1 (b)}
\end{equation}
This remarkably simple result is the one which we sought. Note that, apart
from the norm, it is only involves $\dot{y}_1$ at the boundaries. In
applications, it will usually be the case that the norm in (\ref{simdetpr})
will cancel with an identical factor coming from the lowest order form of
the Jacobian of the transformation to collective coordinates. Therefore
$\langle y_1 | y_1 \rangle$ need not be calculated. Denoting the determinant
of $L$ with the zero mode extracted by $\det 'L$ and normalising by
$\det \hat{L}$, we finally obtain
\begin{equation}
\label{simfin}
\frac{\det 'L}{\det \hat{L}} =
\frac{\langle y_1 | y_1 \rangle}{\dot{y}_1 (a)\dot{y}_1 (b)}
\frac{\dot{\hat{y}}_1 (b)}{\hat{y}_1 (b)}
\end{equation}

The method we have described to find the regularised form of $\det (M+NY(b))$
is hardly more complicated than finding the unregularised form. The key to
achieving this happy state of affairs was firstly the decision to modify
only the boundary conditions, and secondly, the choice of regularised boundary
conditions which gave simple forms for $M^{(\epsilon )}$ and $N^{(\epsilon )}$.
We shall now show that these choices also allow $\lambda ^{(\epsilon )}$ to be
determined in a very simple and elegant way.

\section{The regularisation of the eigenvalue}

While the regularised form of the determinant could be found by use of
Forman's method, a new technique for calculating the previously vanishing
eigenvalue, $\lambda ^{(\epsilon )}$, has to be developed. It is natural to
attempt to calculate it perturbatively in $\epsilon$, but it is not at all
obvious that a general procedure can be set up. Fortunately, it will turn out
that choosing the regularised boundary conditions in the manner illustrated in
section 2 on a simple example, enables $\lambda ^{(\epsilon )}$ to be found to
lowest order almost without calculation.

Let us begin describing the method where the operator is of the simple form
(\ref{simel}); we will generalise to more complicated operators later in this
section. There is no need to specify the boundary conditions at this stage,
since, as we will see, a useful formula for $\lambda ^{(\epsilon )}$ can be
derived without having to make any choices of boundary conditions. Using the
notation introduced in the last section
\begin{equation}
\label{simeig}
L y^{(\epsilon )}_1 = \lambda ^{(\epsilon )} y^{(\epsilon )}_1
\end{equation}
where $y^{(\epsilon )}_1 (t) \rightarrow y_1 (t)$ and
$\lambda ^{(\epsilon )}\rightarrow 0$ as $\epsilon \rightarrow 0$. From
(\ref{simeig})
\begin{eqnarray}
\int^b_a dt \, y_1 L y^{(\epsilon )}_1
& = & \lambda ^{(\epsilon )} \int^b_a dt \, y_1 \, y^{(\epsilon )}_1
\nonumber \\
& = & \lambda ^{(\epsilon )}\langle y_1 | y_1 \rangle \ ,
\label{siminner}
\end{eqnarray}
to lowest order in $\epsilon$. Integrating by parts gives, again to leading
order,
\begin{equation}
\label{simlamform}
\lambda ^{(\epsilon )} = \frac{\left[ \dot{y}^{(\epsilon )}_1 (t) y_1 (t)
- \dot{y}_1 (t) y^{(\epsilon )}_1 (t)\right]^b_a }{\langle y_1 | y_1 \rangle}
\end{equation}
This result is true for operators of the form (\ref{simel}) with arbitrary
boundary conditions. As an example, suppose we impose the regularised boundary
conditions (\ref{simbc}). Then the eigenfunction $y^{(\epsilon )}_1 (t)$
will satisfy them: $y^{(\epsilon )}_1 (a)=0$,
$y^{(\epsilon )}_1 (b)=\epsilon \dot{y}^{(\epsilon )}_1 (b)$. In addition
$y_1 (a)=y_1 (b)=0$, so that to lowest order
\begin{eqnarray}
\label{simreglam}
\lambda ^{(\epsilon )} & = & -\frac{\dot{y}_1 (b) y^{(\epsilon )}_1 (b)}
{\langle y_1 | y_1 \rangle} \nonumber \\
& = & -\epsilon \frac{\dot{y}_1 (b) \dot{y}_1 (b)}
{\langle y_1 | y_1 \rangle}
\end{eqnarray}
as given in section 2. Note that the $\epsilon$ dependence simply came from
the requirement that $y^{(\epsilon )}_1 (b)=\epsilon \dot{y}_1 (b)$, to lowest
order.

Analogous results to (\ref{simlamform}) hold for more general operators. For
example, suppose that
\begin{equation}
\label{el_1}
L_{ij} = \delta _{ij} \frac{d^2\,}{dt^2} + P_{ij} (t) \ \ \ ;
i,j = 1,\ldots ,r
\end{equation}
where $P(t)$ is a complex matrix, and
suppose that the operator (\ref{el_1}) has
a single zero mode $y_{i,1}$, that is, $\sum^{r}_{j=1} L_{ij} y_{j,1} = 0$.
In matrix notation, the zero mode is the column vector
$\underline{y}_1 = (y_{1,1},\ldots ,y_{r,1})^T$. Let
$\underline{y}^{(\epsilon )}_1 (t)$ be the corresponding eigenfunction of the
regularised problem with eigenvalue $\lambda ^{(\epsilon )}$. Then to lowest
order
\begin{equation}
\label{inner_1}
\int^b_a dt \, \sum_{i,j} y^{*}_{i,1} L_{ij} y^{(\epsilon )}_{j,1}
= \lambda ^{(\epsilon )}\sum_{i} \langle y_{i,1} | y_{i,1} \rangle
\end{equation}
where now
\begin{equation}
\label{norm_1}
\langle y_1 | y_1 \rangle \equiv
\sum_{i} \langle y_{i,1} | y_{i,1} \rangle =
\int^b_a dt \, \sum_{i} |y_{i,1}(t)|^2
\end{equation}
Integrating the left-hand-side of (\ref{inner_1}) by parts gives the leading
order result
\begin{equation}
\label{lamform_1}
\lambda ^{(\epsilon )} = \frac{ \sum^{r}_{i=1}\left[ y^{*}_{i,1}(t)
\dot{y}^{(\epsilon )}_{i,1}(t) -
\dot{y}^{*}_{i,1}(t) y^{(\epsilon )}_{i,1}(t) \right]^b_a }
{\langle y_1 | y_1 \rangle } + \frac{\int^b_a dt \, \sum_{i,j}
y^{*}_{i,1}(t)\{ P_{ij} - P^{*}_{ji} \} y^{(\epsilon )}_{j,1}(t)}
{\langle y_1 | y_1 \rangle }
\end{equation}
In most cases of interest to us $L$ will be formally self-adjoint, and so the
second term in (\ref{lamform_1}) will vanish. The self-adjoint nature of $L$ is
expected from its origin as the second functional derivative of the action in
the path integral with respect to the fields:
\begin{equation}
\label{dtwos}
L(t ,t' )_{ij} = \frac{\delta ^2 S}{\delta u^{*}_i (t) \delta u_j (t' )}
\end{equation}

The most general operator which we will study in this paper takes the form
\begin{equation}
\label{el_2}
L_{ij} = [P_{0}(t)]_{ij}\frac{d^2\,}{dt^2} + [P_{1}(t)]_{ij}\frac{d\,}{dt}
+ [P_{2}(t)]_{ij}
\end{equation}
where $P_0 (t)$, $P_1 (t)$ and $P_2 (t)$ are complex $r \times r$ matrices.
We begin by making the transformation
\begin{equation}
\label{pP}
p_{ij}(t) = \exp \left\{ \frac{1}{2} \int^t dt \, (P_0 )^{-1}
(P_1 ) \right\}_{ij} ,\ \ \
P_{ij}(t) = \left[ p\,(P_0 )^{-1} (P_2 ) \, (p)^{-1} \right]_{ij}
- \left[ \ddot{p} \,(p)^{-1} \right]_{ij}
\end{equation}
so that
\begin{equation}
\label{elij}
L_{ij} = (P_0 )_{ik}\, (p^{-1})_{kl} {\cal L}_{lm} \, (p)_{mj}
\end{equation}
where
\begin{equation}
\label{call}
{\cal L}_{ij} = \delta _{ij} \frac{d^2\, }{dt^2} + P_{ij} (t)
\end{equation}
Now if $\hat{L}$ is such that $\hat{P}_0 (t) = P_{0} (t)$, then
\begin{equation}
\label{eqratio}
\frac{\det L}{\det \hat{L}} = \frac{\det {\cal L}}{\det \hat{\cal L}}
\end{equation}
where $\hat{\cal L}$ is as in (\ref{call}), but with $P$ replaced by
$\hat{P}$. Therefore the problem has been reduced to that considered
earlier in this section (see (\ref{el_1}) {\it et seq}). In fact, as
regards determining the ratio of the determinants, Forman gives a general
expression for the left-hand-side of (\ref{eqratio}) (see next section),
and so there is no need to implement the transformation (\ref{pP}). To
find the eigenvalue $\lambda ^{(\epsilon )}$, however, this transformation
is useful. It is easy to see that ${\cal L}$ has a zero mode if, and only if,
$L$ does, and that, in particular, if $y_{i}(t)$ is an eigenfunction of
$L_{ij}$ with zero eigenvalue, then $z_{i}(t) = \sum_{j} p_{ij}(t) y_{j}(t)$
is an eigenfunction of ${\cal L}_{ij}$ with zero eigenvalue. The results
(\ref{inner_1}-\ref{lamform_1}) now hold, but with $L$ and $y$ replaced by
${\cal L}$ and $z$, respectively.
As in all of the examples discussed in this section, a
judicious choice for the boundary conditions on the regularised eigenfunction
$\underline{y}^{(\epsilon )}_1$ will yield an explicit regularised form for
$\lambda ^{(\epsilon )}$ with the minimum of calculational effort.

\section{General procedure}

There are two aspects to our approach to the calculation of
$\det 'L/\det \hat{L}$. One is the operation of finding
$\lambda ^{(\epsilon )}$ to leading order, which was explored for the general
case in the last section. The other aspect concerns the application of Forman's
method for the calculation of $\det L/\det \hat{L}$, but with the regularised
boundary matrices $M^{(\epsilon )}$ and $N^{(\epsilon )}$. This was illustrated
with a simple example in section 2; in this section we discuss Forman's method
in more detail and explain how to apply it to the general operator
(\ref{el_2}).
We end the section with a summary of the general procedure which we have
developed in this paper.

We suppose, following Forman \cite{For}, that the boundary conditions on
(\ref{el_2}) may be expressed as
\begin{eqnarray}
\label{defgenmn}
M\left[\matrix{\underline{u}(a)\cr
               \underline{\dot{u}}(a)\cr}\right] \ \ +
N\left[\matrix{\underline{u}(b)\cr
               \underline{\dot{u}}(b)\cr}\right] \ \ =
\left[\matrix{\underline{0}\cr
              \underline{0}\cr}\right]
\end{eqnarray}
where $M$ and $N$ are $2r \times 2r$ matrices. This equation is simply the
$r$-dimensional analogue of (\ref{defmn}). So for instance, if the boundary
conditions are $\underline{u}(a)=\underline{u}(b)=0$, then
\begin{eqnarray}
\label{mn}
M=\left[\matrix{I_{r}&0\cr
                0&0\cr}\right], \ \ \
N=\left[\matrix{0&0\cr
                I_{r}&0\cr}\right].
\end{eqnarray}
where $I_r$ is the $r \times r$ identity matrix.

Now suppose that $h_i (t); i=1,\ldots , r$, is a solution of the homogeneous
differential equation $\sum_j L_{ij} h_j = 0$, and define the $2r \times 2r$
matrix $Y(t)$, which describes the evolution of a solution and its first
derivative with respect to $t$, by
\begin{eqnarray}
\label{defwhy}
\left[\matrix{\underline{h}(t)\cr
              \underline{\dot{h}}(t)\cr}\right] \ \ =
Y(t)\left[\matrix{\underline{h}(a)\cr
              \underline{\dot{h}}(a)\cr}\right]
\end{eqnarray}
If $\underline{y}_1 (t), \underline{y}_2 (t), ... , \underline{y}_{2r}(t)$,
are $2r$ solutions of $Lh=0$, (\ref{defwhy}) will apply to each
solution separately, that is, $H(t)=Y(t)H(a)$, where
\begin{eqnarray}
\label{hoft_r}
H(t)=\left[\matrix{\underline{y}_1 (t)&\underline{y}_2 (t)&\ldots&
                   \underline{y}_{2r}(t)\cr
                   \underline{\dot{y}}_1 (t)&\underline{\dot{y}}_2 (t)&\ldots&
                   \underline{\dot{y}}_{2r}(t)\cr}\right]
\end{eqnarray}
So, in particular, $H(b)=Y(b)H(a)$, or, if the solutions are independent so
that $\det H \ne 0$,
\begin{equation}
\label{yofb}
Y(b) = H(b)H^{-1}(a)
\end{equation}
This explains the second construction (labelled (ii)) in section 2.

The formula for the ratio of determinants for operators of the type
(\ref{el_2}) is \cite{For}
\begin{equation}
\label{res}
\frac{\det L}{\det \hat{L}} = \frac{\exp \left( \frac{1}{2} \int^b_a dt \,
tr P_1 (t) P_0^{-1}(t) \right) \, \det (M+NY(b))}
{\exp \left( \frac{1}{2} \int^b_a dt \, tr \hat{P}_1 (t)
P_0^{-1}(t) \right) \, \det (M+N\hat{Y}(b))}
\end{equation}
For this result to be applicable, the matrices $P_1 (t)$ and $\hat{P}_1 (t)$,
and also $P_2 (t)$ and $\hat{P}_2 (t)$ need not be equal, however the matrix
$P_0 (t)$, multiplying the second derivative, must be the same for both
operators. In most applications $L$ will be normalised by a $\hat{L}$ which
has a different, and simpler, matrix $P_2$, but is otherwise the same. In
these situations $\hat{P}_1 = P_1$, the exponential factors in
(\ref{res}) cancel out, and the simple formula given by (\ref{simres})
holds (except, of course, that $M, N$ and $Y(b)$ are now $2r \times 2r$, not
$2 \times 2$, matrices). We also note that, although the formula seems to
be asymmetric with respect to the two points $a$ and $b$, one could just as
well define a matrix $\tilde{Y}(t)$ by
\begin{eqnarray}
\label{defwhyhat}
\left[\matrix{\underline{h}(t)\cr
              \underline{\dot{h}}(t)\cr}\right] \ \ =
\tilde{Y}(t)\left[\matrix{\underline{h}(b)\cr
              \underline{\dot{h}}(b)\cr}\right]
\end{eqnarray}
so that $H(a)=\tilde{Y}(a)H(b)$. Then
$\det (M+NY(b))=\det (N+M\tilde{Y}(a))$. Therefore, alternative formulae to
(\ref{simres}) and (\ref{res}) exist, with $M$ and $N$ interchanged and $Y(b)$
replaced by $\tilde{Y}(a)$.

All of the formalism discussed so far in this section also applies to the
problem with regularised boundary conditions --- the only difference is that
$M$ and $N$ are replaced by $M^{(\epsilon )}$ and $N^{(\epsilon )}$
respectively. We are now in a position to summarise the whole procedure:
\begin{enumerate}
\item Modify the boundary conditions of the original problem by a small
amount $(\epsilon )$, so that $\underline{y}_1 (t)$ is no longer a zero
mode. Let $\underline{y}^{(\epsilon )}_1 (t)$ be the eigenfunction of the
new problem with an eigenvalue $\lambda ^{(\epsilon )}$ which tends to zero
as $\epsilon \rightarrow 0$. Express the modified boundary conditions in the
form (\ref{defgenmn}) so that they are characterised by two matrices
$M^{(\epsilon )}$ and $N^{(\epsilon )}$.
\item Calculate $Y(b)=H(b)H^{-1}(a)$, where $H(t)$ is given by (\ref{hoft_r}).
\item Calculate $\det (M^{(\epsilon )}+N^{(\epsilon )}Y(b))$.
\item Calculate $\lambda ^{(\epsilon )}$ from (\ref{lamform_1}).
\item Hence determine
\begin{equation}
\label{detpr}
\lim_{\epsilon \rightarrow 0}
\frac{\det (M^{(\epsilon )}+N^{(\epsilon )}Y(b))}{\lambda ^{(\epsilon )}}
\end{equation}
\item Calculate the denominator factor $\det (M+N\hat{Y}(b))$
\item The ratio of the results of the last two steps gives
$\det 'L/\det \hat{L}$.
\end{enumerate}

We will now study various specific examples where this procedure is applied.

\section{Specific examples}

The algorithm given at the end of the last section gives a method for
determining the ratio $\det 'L/\det \hat{L}$. In this section we will
give explicit results for a few examples with commonly met boundary
conditions and also discuss one example in some detail to show how the method
we have developed works in practice. We will only give results for the
quantity given by (\ref{detpr}), since the final result is found by
normalising this by $\det \hat{L}$, which can be found from the formulae
given in, for example, Forman's paper \cite{For}.

For simplicity we only consider the single component ($r = 1$) case where the
operator has the form (\ref{simel}), for a variety of boundary conditions.
\begin{enumerate}
\item[(a)] With the boundary conditions $Au(a) + B\dot{u}(a) = 0; \ Cu(b)
+ D\dot{u}(b) = 0$,
\begin{equation}
\label{resulta}
\frac{\det 'L}{\langle y_1 | y_1 \rangle } = \left\{ \begin{array}{ll}
+ AC/\dot{y}_1 (a) \dot{y}_1 (b), & \mbox{\ if $A, C \ne 0$} \\
- BC/ y_1 (a) \dot{y}_1 (b), & \mbox{\ if $B, C \ne 0$} \\
- AD/ \dot{y}_1 (a) y_1 (b), & \mbox{\ if $A, D \ne 0$} \\
+ BD/ y_1 (a) y_1 (b), & \mbox{\ if $B, D \ne 0$}
\end{array} \right.
\end{equation}
If all four constants $A, B, C, D$ are non-zero it is easy to see that all
four expressions are equivalent. Similarly, if only three of the constants
are non-zero, then the two applicable expressions are equivalent. If only
two constants are non-zero, one involved in the boundary condition at $a$,
and the other at $b$, then only one of the above applies. The simple example
given in section 2 falls into this class: the boundary conditions there
correspond to $A=1, B=0, C=1, D=0$, and in this case (\ref{resulta}) reduces
to (\ref{simdetpr}).

\medskip

\item[(b)] With periodic boundary conditions $u(a) = u(b); \ \dot{u}(a) =
\dot{u}(b)$,
\begin{equation}
\label{resultb}
\frac{\det 'L}{\langle y_1 | y_1 \rangle } = \frac{ y_2 (b) - y_2 (a) }
{ y_1 (a) \, \det H(a)}
\end{equation}
where $\det H(a) = \dot{y}_2 (a) y_1 (a) - \dot{y}_1 (a) y_2 (a)$ is the
Wronskian.

\medskip

\item[(c)] With anti-periodic boundary conditions $u(a) = - u(b); \
\dot{u}(a) = - \dot{u}(b)$,
\begin{equation}
\label{resultc}
\frac{\det 'L}{\langle y_1 | y_1 \rangle } = - \frac{ y_2 (b) + y_2 (a) }
{ y_1 (a) \, \det H(a)}
\end{equation}

\end{enumerate}

As an example of the application of these results, we use one of the most well
known situations where instantons exist: imaginary time quantum mechanics
with a potential $V(x) = \frac{1}{2} x^2 - \frac{1}{4} x^4$ \cite{blz}. As
shown in the appendix, this problem leads one to consider operators of the
form
\begin{equation}
\label{xfourel}
L = - \frac{d^2\ }{dt^2} + 1 - 3\beta ^2 {\rm dn}^2 (u | m),
\end{equation}
where dn is an elliptic function \cite{as}, $u = \beta (t - t_0 )/\sqrt{2}$
and $\beta = 2(1 - m)/(2 - m)$. The constants $t_0$ and $m$ originate from the
integration of the second order ordinary differential equation which is
satisfied by the instanton. The parameter $t_0$ reflects the breaking of the
time-translational invariance of the original theory and $m$ is related to the
energy of the classical particle in the mechanical analogy. The spectral
properties of the system can be studied by imposing periodic boundary
conditions on the path-integral \cite{schul}, which dictates that we use
(\ref{resultb}) to find the required functional determinant. A straightforward
calculation, outlined in the appendix, yields
\begin{equation}
\label{xfourres}
\frac{\det'L}{\langle y_1 | y_1 \rangle } = - \frac{2 (2 - m)^{7/2}}{m^2}
\left[ \frac{K(m)}{2 - m} - \frac{E(m)}{2(1 - m)} \right]
\end{equation}
where $K(m)$ and $E(m)$ are the complete elliptic integrals of the first and
second kind respectively.

This result simplifies considerably in the limit where the energy of the
particle in the mechanical energy is zero and consequently the period of the
instanton, $T$, becomes infinite. In the appendix it is shown that the
asymptotic forms of (\ref{xfourres}), $\det (M + N\hat{Y}(b))$ and
$\langle y_1 | y_1 \rangle$, for $T$ large are, respectively, $e^T /16$,
$- e^T$ and $\frac{4}{3}$. Combining all of these results gives
\begin{equation}
\label{meqone_4}
\lim_{T \rightarrow \infty} \frac{\det'L}{\det\hat{L}}
= - \frac{1}{12}
\end{equation}
This is in agreement with previous calculations (e.g. eqn. (29) of Ref.
\cite{blz}). It also illustrates the extra complication that may occur if the
range ($a$, $b$) is infinite. In these cases the numerator (\ref{detpr}) and
the denominator $\det (M + N\hat{Y}(b))$ may separately diverge as
$T \equiv (b - a) \rightarrow \infty$. One can avoid these divergences in
various ways, but the most obvious way to proceed in these cases is to use $T$
as a regulator and to perform all calculations with $T$ large, but finite,
cancelling out the potential divergences between numerator and denominator
before taking the $T \rightarrow \infty$ limit.

\section{Conclusions}

In this paper we have developed a simple and effective way of regularising
operators which have zero modes. The method allows the functional determinants
for these kinds of operators, with the zero modes extracted, to be calculated.
The main advantage of the method, and the reason for its power, is that it
leaves much of the structure of the unregularised problem intact. This means
that much of the formalism originally developed in this case can be taken over
with very little change. The approach which we have adopted has not emphasised
rigor; it would be very interesting to put this work on a rigorous footing.
In particular, we have not proved
that the results are independent of the precise method of regularisation
adopted. Until this is done, the results obtained using our method, especially
for $r>1$, should be treated with caution. On the other hand, we have also
kept the number of examples of the application of the technique to a minimum,
preferring instead to give a clear and explicit discussion of the methodology.

Although we have tried to be quite general, describing most aspects of
the formalism which may arise in practice, there are, inevitably, situations
that have not been covered. One is the case where there is more than one zero
mode present --- an example is the model studied in \cite{tsg}.
The procedure in cases such as this is a simple extension of our previous
discussion: regularising parameters $\epsilon_{1}, \epsilon_{2}, \ldots$ are
introduced for every broken symmetry, and hence for every zero mode. This
symmetry may be external (spatial or temporal) or internal (global or local).
The boundary
conditions are then modified along the directions of breaking by an amount
$\epsilon_{\alpha}$, and the prescription given in section IV followed.

Our motivation for carrying out this work has been the increasing need to
evaluate determinants
of this kind in many areas of the physical sciences. We hope that the ideas
presented here are sufficiently straightforward and easily implemented
that they will find wide application.

\acknowledgements

We wish to thank Paul Goldbart for useful discussions. AJM wishes to thank the
Department of Physics at the University of Illinois at Urbana-Champaign for
hospitality. This work was supported in part by EPSRC grant GR/H40150 (AJM) and
by NSF grant No. DMR-89-20538 (MBT). One of us (MBT), gratefully acknowledges
support from a U.S. Department of Education GAANN Fellowship.

\appendix
\section*{}

In this appendix we give details of the calculation of the functional
determinant of (\ref{xfourel}). The motivation for studying an operator of this
type is that it arises in the investigation of fluctuations about the instanton
in one-dimensional quantum mechanics with the potential
$V(x) = \frac{1}{2} x^2 - \frac{1}{4} x^4$ \cite{blz}. The instanton satisfies
the equation $- \ddot{x} + V'(x) = 0$, which may be integrated once to give
$\frac{1}{2}\dot{x}^2 - V(x) = E$, where $E$ is a constant. Solutions to this
equation are those of a classical particle of unit mass and energy $E$ moving
in the potential $-V(x)$. Bounded motion is allowed for $E < 0$, corresponding
to the existence of real instantons.

Let the values of $x$ at which the particle in this mechanical analogy has
zero velocity be denoted by $\alpha$ and $\beta$ ($0 < \alpha < \beta$). Then
$- V (\alpha ) = - V( \beta ) = E$ which implies $\alpha ^2 + \beta ^2 = 2$
and $E = -\alpha ^2 \beta ^2 /4$. The once integrated equation of motion now
reads
\begin{equation}
\label{firstorder}
\left( \frac{dx}{dt} \right)^2 = \frac{1}{2} (x^2 - \alpha ^2 )
(\beta ^2 - x^2 )
\end{equation}
\begin{equation}
\label{elliint}
\Rightarrow \ \int^{x}_{\beta} \frac{dx}
{\sqrt{(x^2 - \alpha ^2 )(\beta ^2 - x^2 )}}
= - \frac{1}{\sqrt{2}} \, (t - t_0 )
\end{equation}
where $t_0$ is the time at which the particle was at $x = \beta$. This may be
integrated in terms of elliptic functions \cite{as}:
\begin{equation}
\label{xc}
x_c (t; t_{0}, m) = \beta \, {\rm dn}(u|m)
\end{equation}
where $u = \beta (t - t_0 )/\sqrt{2}$ and $m = 1 - (\alpha ^2 /\beta ^2 )$.
The subscript `$c$' denotes ``classical" and simply indicates that this is a
solution of the classical equation of motion $\delta S/\delta x(t) = 0$,
where $S[x]=\int_{a}^{b}dt\left[ \frac{1}{2}{\dot x}^{2}+V(x)\right]$ is
the action. The
physical significance of the integration constant $t_0$ is clear: since the
particle can start at any $x \ (\alpha \le x \le \beta )$, the time at which
it reaches $\beta$ (defined to be $t_0$) is arbitrary. The constant $m$, on
the other hand, is directly related to the energy of the particle, since
$E = - (1 - m )/2 (2 - m)^2$. An alternative to $m$, which also specifies the
energy of the particle, is the period $T$ defined by
\begin{eqnarray}
\frac{T}{2} \, \frac{1}{\sqrt{2}}
& = & \int^{\beta}_{\alpha} \frac{dx}
{\sqrt{(x^2 - \alpha ^2 )(\beta ^2 - x^2 )}} \nonumber \\
& = & \beta ^{-1}\, \int^{\pi /2}_{0} \frac{d\theta}
{\sqrt{ 1 - m\sin ^2 \theta }}
\label{period_1} \\
& = & \left( \frac{2 - m}{2} \right)^{1/2}\, K(m)
\label{period_2}
\end{eqnarray}
where $K(m)$ is the complete elliptic integral of the first kind \cite{as}.

As explained in the main part of the text, we are interested in evaluating the
expression (\ref{resultb}), and therefore need to determine the values of the
functions $y_1$ and $y_2$ at the endpoints $a$ and $b$. These two functions are
solutions of the homogeneous differential equation $Lh = 0$, where
\begin{equation}
\label{secondvar}
L \delta (t - t') = \left. \frac{\delta ^2 S}
{\delta x(t) \delta x(t')} \right|_{x=x_c} = \left[ - \frac{d^2 \ }{dt^2} + 1
- 3x_c ^2 (t;t_{0}, m) \right] \delta (t - t')
\end{equation}
Using the explicit form for $x_c$ given by (\ref{xc}) we obtain
(\ref{xfourel}). But two independent solutions of $Lh = 0$ can be found by
differentiating $x_c$ with respect to $t_0$ and $m$ \cite{Fox}, so we define
$y_1$ and $y_2$ by:
\begin{equation}
\label{yone}
y_1 (t; t_{0}, m) \equiv \frac{\partial x_c (t; t_{0}, m)}{\partial t_0}
\end{equation}
\begin{equation}
\label{ytwo}
y_2 (t; t_{0}, m) \equiv \frac{\partial x_c (t; t_{0}, m)}{\partial m}
\end{equation}
It is a straightforward exercise in elliptic functions to find from (\ref{xc})
that:
\begin{eqnarray}
y_1 (t; t_{0}, m) & = & \frac{m\beta ^2}{\sqrt{2}}\, {\rm sn}(u|m)
\, {\rm cn}(u|m) \label{exyone} \\
\dot{y}_1 (t; t_{0}, m) & = & \frac{m\beta ^3}{2}\, {\rm dn}(u|m)
\, \left\{ {\rm cn}^2 (u|m) - {\rm sn}^2 (u|m) \right\} \label{exyonedot}
\end{eqnarray}

One can now check using (\ref{period_2}) that $y_1 (a) = y_1 (b)$ and that
$\dot{y}_1 (a) = \dot{y}_1 (b)$ for any initial and final times satisfying
$b - a = T$. Therefore, since $y_1$ is a solution of $Lh = 0$ satisfying
the correct boundary conditions, it is the zero mode for this problem, as
expected.

A slightly longer calculation gives
\begin{eqnarray}
\label{exytwo}
y_2 (t; t_{0}, m) & = & \frac{d\beta}{dm}\, {\rm dn}(u|m) - \left\{ u m
\frac{d\beta}{dm} - \frac{\beta \, E(u|m)}{2(1 - m)}
+ \frac{\beta u}{2} \right\}{\rm sn}(u|m){\rm cn}(u|m) \nonumber \\
& - & \frac{\beta {\rm sn}^2 (u|m)\, {\rm dn}(u|m)}{2(1 - m)}
\end{eqnarray}
where $E(u|m)$ is the elliptic integral of the second kind. Using the
periodicity of the elliptic functions
\begin{eqnarray}
\label{inter}
\frac{y_2 (b) - y_2 (a)}{y_1 (a)} & = & -2 \left\{ K(m)m\frac{d\beta}{dm}
- \frac{\beta E(m)}{2(1 - m)} + \frac{\beta}{2}\, K(m) \right\}
\left\{ \frac{\beta ^2 m}{\sqrt{2}} \right\}^{-1} \nonumber \\
& = & - 2 \, \frac{(2 - m)^{1/2}}{m}\left\{ \frac{K(m)}{2 - m}
- \frac{E(m)}{2(1 - m)} \right\}
\end{eqnarray}
since $\beta ^2 = 2(2 - m)^{-1}$.
Here $E(m)$ is the complete elliptic integral of the second kind.

The Wronskian $\det H(t)$ is a constant, and so can be calculated for any
convenient $t$. Choosing $t = t_0$, which
implies $u=0$ and so $y_{1}(t_{0}) = 0$,
\begin{eqnarray}
\det H(t) & = & \dot{y}_2 (t_0 )\, y_1 (t_0 ) - \dot{y}_1 (t_0 )\,
y_2 (t_0 ) \nonumber \\
& = & - \left( \frac{\beta ^3 m}{2} \right) \left( \frac{d\beta}{dm} \right)
\nonumber \\
& = & - \frac{m}{(2 - m )^3}
\label{exwron}
\end{eqnarray}
Substituting (\ref{inter}) and (\ref{exwron}) into (\ref{resultb}), and taking
into account the extra minus sign which comes about because the operator
(\ref{xfourel}) is {\it minus} the definition of operators as given in the
text, gives (\ref{xfourres}).

Following the discussion of the most natural form for $\hat{L}$ given in
section II, we take it to be the second functional derivative of the action for
the harmonic oscillator with the potential $\hat{V}(x) = \frac{1}{2} x^2$.
Then
\begin{equation}
\label{exlhat}
\hat{L} = - \frac{d^2\ }{dt^2} + 1
\end{equation}
Choosing $\hat{y}_1 (t) = e^t$ and $\hat{y}_2 (t) = e^{-t}$ to be the two
independent solutions of the homogeneous equation $\hat{L}h = 0$,
\begin{eqnarray}
\label{exyhat}
\hat{Y}(b)=\left[\matrix{\cosh (b-a)&\sinh (b-a)\cr
                   \sinh (b-a)&\cosh (b-a)\cr}\right].
\end{eqnarray}
Using the same periodic boundary conditions which gave (\ref{resultb}),
\begin{eqnarray}
\label{exdetlhat}
\det (M + N\hat{Y}(b) ) & = & 2 - \hat{Y}_{11}(b) - \hat{Y}_{22}(b)
\nonumber \\
& = & 2(1 - \cosh T ) \label{dethat_1} \\
& = & - 4\sinh ^2 ([2 - m]^{1/2}\, K(m)) \label{dethat_2}
\end{eqnarray}
Dividing (\ref{xfourres}) by (\ref{dethat_2}) gives the required expression for
\begin{equation}
\label{finalres}
\frac{1}{\langle y_1 | y_1 \rangle} \, \frac{\det 'L}{\det\hat{L}}
\end{equation}

As a check on the results let us look at the limit $E \rightarrow 0_{-}$, i.e.
$m \rightarrow 1$ or $T \rightarrow \infty$. In this case
$K(m) \sim \frac{1}{2}\ln (1 - m)$, which from (\ref{period_2}) gives
$m \sim 1 - 16\, e^{-T}$. Using $E(m) \rightarrow 1$ as $m \rightarrow 1$, we
have
\begin{equation}
\label{larget_1}
\frac{\det 'L}{\langle y_1 | y_1 \rangle } \sim \frac{e^T}{16}, \ \
{\rm as \ }T \rightarrow \infty
\end{equation}
Since from (\ref{dethat_1}), $\det\hat{L} \sim - e^T$ as $T \rightarrow
\infty$,
\begin{equation}
\label{larget_2}
\lim_{T \rightarrow \infty} \frac{1}{\langle y_1 | y_1 \rangle}\,
\frac{\det 'L}{\det\hat{L}} = - \frac{1}{16}
\end{equation}

The sign is the expected one: the zero mode which has been extracted, $y_1$,
has a single node, which leads us to deduce that $L$ has only one
eigenfunction with a negative eigenvalue; all the other eigenvalues are
non-negative. The signs of (\ref{dethat_2}) and (\ref{larget_1}) are not those
that we might naively expect, but these signs have no meaning separately ---
both the magnitude and sign of these terms can be changed at will by the
replacement $M \rightarrow \lambda M, N \rightarrow \lambda N$, where
$\lambda$ is any real number.

The ratio (\ref{larget_2}) agrees with the calculation of Ref. \cite{blz}. To
see this we note that $\alpha \rightarrow 0, \beta \rightarrow \sqrt{2}$ as
$m \rightarrow 1$, hence the instanton becomes
\begin{eqnarray}
x_c (t; t_{0}, m=1) & = & \sqrt{2}\, {\rm sech}(t - t_{0}) \label{meqone_1} \\
\Rightarrow y_{1}(t; t_{0}, m=1) & = & \sqrt{2}\, {\rm sech}(t - t_{0})
\tanh (t - t_{0}) \label{meqone_2} \\
\Rightarrow \lim_{T \rightarrow \infty} \langle y_1 | y_1 \rangle
& = & \frac{4}{3} \label{meqone_3}
\end{eqnarray}
Combining (\ref{larget_2}) and (\ref{meqone_3}) gives (\ref{meqone_4}), as
required.

\end{document}